\documentclass[aps,pra,reprint,noeprint]{revtex4-1} 
\usepackage{bm, braket,amsmath,mathtools,comment,longtable}
\usepackage[T1]{fontenc}
\usepackage{graphicx,color}
\usepackage{hyperref}

\begin{document}
\title{Minimally entangled typical thermal states algorithm with Trotter gates}
\author{Shimpei Goto}
\email[]{goto@phys.kindai.ac.jp}
\author{Ippei Danshita}
\email[]{danshita@phys.kindai.ac.jp}
\affiliation{Department of Physics, Kindai University, Higashi-Osaka city, Osaka, Japan}
\date{\today}
\begin{abstract}
  We improve the efficiency of the minimally entangled typical thermal states (METTS) algorithm without breaking the Abelian symmetries.
  By adding the operation of Trotter gates that respects the Abelian symmetries to the METTS algorithm, 
  we find that a correlation between successive states in Markov-chain Monte Carlo sampling decreases by orders of magnitude.
  We measure the performance of the improved METTS algorithm through the simulations of the canonical ensemble of the Bose-Hubbard model and confirm that the reduction of the autocorrelation leads to the reduction of computation time.
  We show that our protocol using the operation of Trotter gates is effective also for the simulations of the grand canonical ensemble.
\end{abstract}
\maketitle
\section{Introduction\label{sec:introduction}}
One-dimensional (1D) systems are exceptional in the sense that it is tractable with classical computers to simulate their quantum many-body physics.
The entanglement of a pure state in a quantum system is often quantified using the entanglement entropy of a subsystem that consists of a left (or right) half of the system.
It is widely known that in 1D the entanglement of a ground state is not extensive, i.e., as a function of the system size it is constant in a system when the energy of the lowest excitation is gapped or increases only logarithmically in a gapless system~\cite{hastings_area_2007,hastings_entropy_2007,eisert_textitcolloquium_2010}.
Thanks to the low entanglement, a ground state of a generic 1D quantum system 
can be represented efficiently by matrix product states (MPS)~\cite{schollwock_density-matrix_2011} and can be numerically obtained via some smart optimization algorithms, including the density matrix renormalization group~\cite{white_density_1992,white_density-matrix_1993} and imaginary-time evolution using the time-evolving block decimation (TEBD)~\cite{vidal_efficient_2003,vidal_efficient_2004,daley_time-dependent_2004,white_real-time_2004}.
Moreover, the MPS representation has been applied for analyzing real-time evolution of a pure state~\cite{vidal_efficient_2003,vidal_efficient_2004,daley_time-dependent_2004,white_real-time_2004,garcia-ripoll_time_2006,paeckel_time-evolution_2019-1} and a system at finite temperature described by a mixed state~\cite{zwolak_mixed-state_2004,verstraete_matrix_2004,feiguin_finite-temperature_2005,white_minimally_2009,stoudenmire_minimally_2010,iwaki_thermal_2020}.
In general, however, since highly excited states with extensive entanglement entropies are involved in such MPS simulations of real-time dynamics and finite-temperature systems, one needs some ways to circumvent the infeasibility of directly representing such states in terms of MPS.\@

For the MPS simulations of finite temperature systems, 
there exist two major approaches: the purification method in an enlarged Hilbert space~\cite{verstraete_matrix_2004,feiguin_finite-temperature_2005} or the sampling method known as the minimally entangled typical thermal states (METTS) algorithm~\cite{white_minimally_2009,stoudenmire_minimally_2010}.
In the former method, a mixed state is represented as a pure state in the enlarged Hilbert space.
Because of the enlargement, the computational cost of the purification method increases polynomially but quickly as the matrix dimensions of MPS, namely bond dimensions, increase. 
Since the bond dimensions increase for decreasing the temperature, the approach is efficient at high temperatures.
Besides, a variant specialized for low temperature systems has also been proposed~\cite{goto_cooling_2017}.
At intermediate temperatures, the purification method often requires a tremendous computational cost.

In the METTS algorithm, a mixed state is represented via a Markov-chain Monte Carlo (MCMC) sampling of pure states in the original Hilbert space, and one does not have to treat the enlarged Hilbert space.
Moreover, the most computationally expensive operation is the imaginary-time evolution of MPS in the original Hilbert space, and thus the METTS algorithm does not suffer from a tremendous cost in systems where the ground state algorithm by imaginary-time evolution works well.
Thanks to these advantages, the METTS algorithm is expected to be able to access a broad temperature range and seems superior to the purification method in numerical efficiency.
Nonetheless, thus far, there have been more applications of the purification method~\cite{barthel_spectral_2009,endres_observation_2011,karrasch_finite-temperature_2012,karrasch_nonequilibrium_2013,karrasch_drude_2013,barthel_precise_2013,lake_multispinon_2013,huang_scaling_2013,gori_finite-temperature_2016,bruognolo_dynamic_2016,bruognolo_matrix_2017,goto_cooling_2017,nocera_finite-temperature_2018} than those of the METTS algorithm~\cite{yao_thermalization_2012,alvarez_implementation_2012,bonnes_light-cone_2014,lacki_dynamics_2015,bruognolo_matrix_2017,goto_quasiexact_2019}.
Moreover, it has been reported that the METTS algorithm is less efficient than the purification method because of statistical errors induced by the sampling~\cite{binder_minimally_2015}.

One of the major roots of the inefficiency is the autocorrelation of successive samples which reduces the effective number of samples~\cite{white_minimally_2009,stoudenmire_minimally_2010,binder_minimally_2015}.
There exists a remedy to reduce the autocorrelation in the METTS algorithm~\cite{white_minimally_2009,stoudenmire_minimally_2010}. 
However, this remedy changes the total magnetization or particle numbers during simulations, i.e., it disregards the Abelian symmetries.
In a viewpoint of statistical mechanics, the unfixed quantum numbers only mean that statistical ensemble is the ground canonical ensemble and do not bring a significant problem as long as a system size is large enough.
On the contrary, the disregard of the Abelian symmetries dramatically spoils the numerical efficiency~\cite{schollwock_density-matrix_2011,lacki_dynamics_2015}.

In our previous work~\cite{goto_quasiexact_2019}, in order to simulate a finite-temperature dynamics of the Kondo model~\cite{kondo_resistance_1964} in a quasi-exact manner, we have improved the METTS algorithm by introducing a unitary operation represented as a series of Trotter gates. 
This approach is a variant of the symmetric METTS algorithm developed by \textcite{binder_symmetric_2017} and is more flexible thanks to the controllability of the time step and the Hamiltonian constructing the Trotter gates.
Our approach has allowed us to access the logarithmic temperature dependence of the transport, which is a characteristic feature of the Kondo effects~\cite{kondo_resistance_1964,goto_quasiexact_2019}.

In this paper, we present more extensive and quantitative analyses of the improved METTS algorithm.
We discuss how to utilize the flexibility of the unitary operation composed by Trotter gates in order to remove the autocorrelation problem in systems with large gaps such as a strongly interacting Bose-Hubbard model.
We also find that our protocol can be combined with the hybrid approach of the purification method and the METTS algorithm, which has been developed very recently~\cite{chung_minimally_2019,chen_hybrid_2020}.
Taking the 1D Bose-Hubbard model as a specific example, we present benchmark tests for the approaches developed in this work.

The rest of the paper is organized as follows: In Sec.~\ref{sec:methods}, we briefly review the METTS algorithm and introduce its generalization including the basis transformation made by the operation of Trotter gates. 
In Sec.~\ref{sec:results}, we present performance tests with the Bose-Hubbard model. 
In Sec.~\ref{sec:summaries}, we summarize the results.

\section{minimally entangled typical thermal states algorithm\label{sec:methods}}
\subsection{Minimally entangled typical thermal states algorithm and its autocorrelation problem}
The objective of the METTS algorithm is to compute the thermal expectation value of an operator \(\hat{O}\) at inverse temperature \( \beta \), which is given by 
\begin{align}
  \begin{aligned}
  \braket{\hat{O}} &= \frac{1}{Z}\mathrm{Tr}\left[\mathrm{e}^{-\beta \hat{H}} \hat{O}\right] \\
  &= \sum_i \frac{\braket{i|\mathrm{e}^{-\beta \hat{H}}|i}}{Z}\frac{\braket{i|\mathrm{e}^{-\frac{\beta}{2}\hat{H}}\hat{O}\mathrm{e}^{-\frac{\beta}{2}\hat{H}}|i}}{\braket{i|\mathrm{e}^{-\beta \hat{H}}|i}},
  \end{aligned}
\end{align}
where the summation takes over an orthonormal basis \( \ket{i}\), \( Z = \sum_i \braket{i|\mathrm{e}^{-\beta\hat{H}}|i}\), and \(\hat{H}\) is the Hamiltonian of the system. 
Since the number of \( \ket{i}\) increases exponentially with system size, it is impossible to perform the summation over all \( \ket{i}\) in large systems. Hence, one has to introduce a sampling method.

The METTS algorithm efficiently performs this sampling utilizing the fact that MPS,
\begin{align}
  \ket{\psi} = \sum_{\bm{\sigma}} \bm{A}^{\sigma_1}_1 \bm{A}^{\sigma_2}_2 \ldots \bm{A}^{\sigma_L}_L \ket{\bm{\sigma}},
\end{align}
can efficiently represent low-entangled states~\cite{hastings_entropy_2007,schollwock_density-matrix_2011}.
Here, \(\sigma_m\) is the index of the local Hilbert space at site \(m\), 
\(L\) is the number of lattice sites,
\(\ket{\bm{\sigma}} = \ket{\sigma_1, \sigma_2, \ldots, \sigma_L}\), 
and \(\sum_{\bm{\sigma}}\) means the summation over all possible configurations of \({\sigma_i}\).
The matrix dimensions of matrices \(\bm{A}^{\sigma_m}_m\) are called bond dimensions.
In the METTS algorithm, one takes classical product states (CPS) as the orthonormal basis~\cite{white_minimally_2009,stoudenmire_minimally_2010}: \( \ket{i} = \ket{\bm{\sigma}}\).
Each CPS state can be represented by MPS with bond dimension \(\chi=1\).

Basic procedure of the METTS algorithm is as follows.
Starting from a certain initial CPS \(\ket{i} = \ket{\bm{\sigma}}\), one calculates a state
\begin{align}
  \label{eq:phi_i}
  \ket{\phi(i)} = \frac{\mathrm{e}^{-\frac{\beta}{2}\hat{H}}\ket{i}}{\sqrt{{\braket{i|\mathrm{e}^{-\beta\hat{H}}|i}}}}
\end{align}
by imaginary-time evolution and the expectation value \(\braket{\phi(i)|\hat{O}|\phi(i)}\).
The state \(\ket{\phi(i)}\) is subsequently projected into a new CPS \(\ket{j}\) with the probability
\begin{align}
  \label{eq:prob_metts}
  p_{i \to j} = |\braket{j|\phi(i)}|^2
\end{align}
and one repeats the imaginary-time evolution and the observation.
This is a MCMC algorithm and its stationary distribution \( \Pi_i\) is given as the left eigenvector with eigenvalue 1 of a transition matrix whose elements are defined by \(p_{i,j} = p_{i \to j}\)~\cite{sokal_monte_1997}.
One can easily confirm that the canonical ensemble
\begin{align}
  \label{eq:static_dist}
  \Pi_i = \frac{\braket{i|\mathrm{e}^{-\beta \hat{H}}|i}}{Z},
\end{align}
is the stationary distribution of the Markov chain generated by the METTS algorithm as follows:
\begin{align}
\begin{aligned}
  \label{eq:calc_stationary}
  \sum_i \Pi_i p_{i, j} &= \sum_i \frac{\braket{i|\mathrm{e}^{-\beta \hat{H}}|i}}{Z} \frac{\braket{i|\mathrm{e}^{-\frac{\beta}{2} \hat{H}}|j}\braket{j|\mathrm{e}^{-\frac{\beta}{2} \hat{H}}|i}}{\braket{i|\mathrm{e}^{-\beta \hat{H}}|i}} \\ 
  &= \frac{1}{Z} \sum_i \braket{j|\mathrm{e}^{-\frac{\beta}{2}\hat{H}}|i}\braket{i|\mathrm{e}^{-\frac{\beta}{2}\hat{H}}|j}  \\
  &= \Pi_j.
\end{aligned}
\end{align}

The METTS algorithm efficiently simulates the thermodynamic properties of spin-rotationally invariant spin systems~\cite{white_minimally_2009,stoudenmire_minimally_2010}.
However, it has been reported that a strong correlation of successive samples arises when the METTS algorithm is applied to the Bose-Hubbard model with particle-number conservation~\cite{lacki_dynamics_2015}.
The existence of such a severe autocorrelation problem can be inferred from Eqs.~\eqref{eq:phi_i} and~\eqref{eq:prob_metts} by taking a small inverse temperature \(\beta \).
At a small \(\beta \), \( \ket{\phi(i)}\) has large overlap with \(\ket{i}\) so that the probability for choosing \(\ket{i}\) again is very high. 
A severe autocorrelation problem also arises when a CPS has large overlap with one of eigenstates of \(\hat{H}\). 
In SU(2)-symmetric spin systems, these problems can be eliminated by changing the spin axis of CPS, e.g., \(\ket{\phi(i)}\) is projected into a CPS with X-axis for odd steps and projected into a CPS with Z-axis for even steps~\cite{white_minimally_2009,stoudenmire_minimally_2010}.
Although a CPS projected into X-axis is a superposition of states with different magnetization in Z-axis, the CPS is in a symmetric sector of magnetization in X-axis and the Hamiltonian conserve the magnetization in X-axis.
In other words, the Abelian symmetry can be utilized.
If this procedure is straightforwardly applied to particle systems, on the contrary, a resulting state is a superposition of states with different numbers of particles and there exists no Abelian symmetry that can be utilized unlike SU(2)-symmetric spin systems.
Hence, numerical simulations become very inefficient.

\subsection{Symmetric bases given by real-time evolution\label{sec:sym_base}}
In order to relax the autocorrelation problem of the METTS algorithm with the Abelian symmetry respected, 
Binder and Barthel have incorporated the use of different symmetric bases for different Monte Carlo steps~\cite{binder_symmetric_2017}.
Here, ``symmetric'' means that states in a symmetric basis lie within a certain subspace, in which the eigenvalue of an operator associated with an Abelian symmetry is fixed.
In this subsection, we generalize their approach and introduce an easy way to obtain various symmetric bases.

A symmetric basis \(\ket{i}_g\) with respect to an operator \(\hat{G}\) is an eigenbasis of \(\hat{G}\) with an eiganvalue \(g\), i.e., \(\hat{G}\ket{i}_g = g\ket{i}_g\) for any \(i\).
Another symmetric basis \(\ket{i^\prime}_g\) with the same eigenvalue can be obtained by a unitary transformation \(\hat{U}\), i.e., \(\ket{i^\prime}_g = \hat{U}\ket{i}_g\).
One can easily confirm that the basis \(\ket{i^\prime}_g\) is also an eigenbasis with the same eigenvalue as long as the unitary operator \(\hat{U}\) commutes with the operator \(\hat{G}\):
\begin{align}
  \hat{G}\ket{i^\prime}_g &= \hat{G}\hat{U}\ket{i}_g \nonumber \\
  &= \hat{U}\hat{G}\ket{i}_g \nonumber \\
  &= g\ket{i^\prime}_g.
\end{align}
By representing a unitary operator as 
\begin{align}
  \hat{U} = \mathrm{e}^{-\mathrm{i}\tau \hat{A}}
\end{align}
with a real parameter \(\tau \) and a hermitian operator \(\hat{A}\), 
the condition for the commutability of \(\hat{U}\) and \(\hat{G}\) can be recast into that of \(\hat{A}\) and \(\hat{G}\). 
In other words, real-time evolution via any Hamiltonian commuting with \(\hat{G}\) gives a new symmetric basis.
Changing the operator \(\hat{A}\) and the parameter \(\tau \), one can obtain many symmetric bases.
This variety of bases is one advantage of our proposed symmetric bases given by real-time evolution.
Another advantage of our protocol for creating symmetric bases is that it is easy to implement.
More specifically, since a time-evolution method such as the TEBD is required in the METTS algorithm for obtaining a state \(\ket{\phi(i)}\) of Eq.~\eqref{eq:phi_i}, the operation \(\ket{i^\prime}_g = \mathrm{e}^{-\mathrm{i}\tau\hat{A}}\ket{i}_g\) can be implemented immediately.
Hereafter, we consider only symmetric bases of a certain operator and drop the subscript \(g\).

As a specific procedure for utilizing the creation protocol of a symmetric basis to reduce the autocorrelation problem, we project an imaginary-time evolved state into a CPS \(\ket{i}\) for even steps of the MCMC sampling and into a transformed state \(\ket{i^\prime} = \hat{U}\ket{i}\) for odd steps.
This procedure does not change the stationary distribution of the Markov chain as shown in the followings.
From Eq.~\eqref{eq:prob_metts}, the transition probability from a CPS \(\ket{i}\) to a transformed state \(\ket{k^\prime}\) in odd steps is given as
\begin{align}
  \label{eq:prob_q}
  \begin{aligned}
  q_{i \to k} & = \frac{\left|\braket{k^\prime|\mathrm{e}^{-\frac{\beta}{2}\hat{H}}|i}\right|^2}{\braket{i|\mathrm{e}^{-\beta \hat{H}}|i}} \\
  &= \frac{\braket{k|\hat{U}^\dagger \mathrm{e}^{-\frac{\beta}{2}\hat{H}}|i}\braket{i|\mathrm{e}^{-\frac{\beta}{2}\hat{H}}\hat{U}|k}}{\braket{i|\mathrm{e}^{-\beta \hat{H}}|i}}.
  \end{aligned}
\end{align}
Similarly, the transition probability from a transformed state \(\ket{k^\prime}\) to a CPS \(\ket{j}\) in even steps is given as
\begin{align}
  \label{eq:prob_r}
  \begin{aligned}
  r_{k \to j} & = \frac{\left|\braket{j|\mathrm{e}^{-\frac{\beta}{2}\hat{H}}|k^\prime}\right|^2}{\braket{k^\prime|\mathrm{e}^{-\beta \hat{H}}|k^\prime}}\\
  &= \frac{\braket{j|\mathrm{e}^{-\frac{\beta}{2}\hat{H}}\hat{U}|k}\braket{k|\hat{U}^\dagger\mathrm{e}^{-\frac{\beta}{2}\hat{H}}|j}}{\braket{k|\hat{U}^\dagger\mathrm{e}^{-\beta \hat{H}}\hat{U}|k}}.
  \end{aligned}
\end{align}
Considering the two steps as one step, the transition probability from a CPS \(\ket{i}\) to a CPS \(\ket{j}\) is given as 
\begin{align}
  \label{eq:nmetts_prob}
  p_{i \to j} = \sum_k q_{i \to k} r_{k \to j}.
\end{align}
One can confirm that the canonical ensemble \(\Pi_i\)~\eqref{eq:static_dist} is the stationary distribution of the Markov chain with this transition probability as follows:
\begin{align}
\begin{aligned}
  \sum_{i}\Pi_i p_{i, j}  &= \frac{1}{Z}\sum_{i, k} \braket{k|\hat{U}^\dagger \mathrm{e}^{-\frac{\beta}{2}\hat{H}}|i}\braket{i|\mathrm{e}^{-\frac{\beta}{2}\hat{H}}\hat{U}|k}r_{k \to j} \\
  & = \frac{1}{Z}\sum_k \braket{j|\mathrm{e}^{-\frac{\beta}{2}\hat{H}}\hat{U}|k}\braket{k|\hat{U}^\dagger\mathrm{e}^{-\frac{\beta}{2}\hat{H}}|j} \\
  & = \Pi_j.
\end{aligned}
\end{align}
On the other hand, the transition probability from a transformed state \(\ket{i^\prime}\) to a transformed state \(\ket{j^\prime}\) is given as
\begin{align}
  p^\prime_{i \to j} = \sum_k r_{i \to k} q_{k \to j}
\end{align}
and one can confirm that the stationary distribution of the Markov chain defined by this transition probability is also the canonical ensemble in \(\ket{i^\prime}\) basis,
\begin{align}
  \Pi^\prime_i = \frac{\braket{i|\hat{U}^\dagger \mathrm{e}^{-\beta \hat{H}}\hat{U}|i}}{Z}.
\end{align}
Thus, the stationary distribution of a CPS \(\ket{i}\) in odd steps is \(\Pi_i\) and that of a transformed state \(\ket{i^\prime}\) in even steps is \(\Pi^\prime_i\).
It should be noted that the thermal expectation values do not depend on bases because of the similarity invariance of the trace.
Hence, samples in both odd and even steps can be used to estimate the thermal expectation values.

If a state \(\hat{U}^\dagger \mathrm{e}^{-\frac{\beta}{2}\hat{H}}\ket{i}\) is the superposition of many CPS with similar weights, the above-mentioned procedure significantly reduces the autocorrelation problems.
This is because the probability \(q_{i \to k}\)~\eqref{eq:prob_q} has also similar values to many \(k\)s so that a correlation of successive samples is small.
However, there is a trade-off.
While more efficient reduction of the autocorrelation requires \(\hat{U}^\dagger \mathrm{e}^{-\frac{\beta}{2}\hat{H}}\ket{i}\) to consist of many CPS, i.e., to be more entangled, manipulations of highly entangled MPS are numerically expensive.
Hence, a controller to adjust the entanglement induced by \(\hat{U}\) is necessary and we introduce the parameter \(\tau \) for this purpose.
In a typical situation, larger \(\tau \) makes the entanglement larger but the computation of a single Monte Carlo step more costly.

As the operator \(\hat{A}\), any choice might be effective as long as an operator respects the Abelian symmetry.
A straightforward choice for \(\hat{A}\) is the Hamiltonian of the system \(\hat{H}\) and this choice is sufficient for most cases.
However, in the case that there are some CPS being overlapped largely with some of the eigenstates of \(\hat{H}\), this straightforward choice might be ineffective because \(\tau \) is required to be rather large for creating a state composed of many CPS\@.
Examples include the Bose-Hubbard model with integer filling and the strong on-site interaction, and the strongly anisotropic Heisenberg model.
In such cases, one should choose another Hamiltonian like that of the free bosons or the isotropic Heisenberg model. 

Since the real-time evolution of MPS is a numerically expensive tasks in general, how to implement the application of an operator \(\hat{U}\) is also important for numerical efficiency.
We here emphasize that what is necessary for conducting our procedure is not real-time evolution via a certain Hamiltonian but a unitary operation that makes the state optimally entangled.
Therefore, the application of the Trotter decomposed operator
\begin{align}
  \label{eq:1stTrotter}
  \hat{U}_\mathrm{T}(\tau) = \mathrm{e}^{-\mathrm{i}\tau \hat{H}_\mathrm{even}}\mathrm{e}^{-\mathrm{i}\tau \hat{H}_\mathrm{odd}},
\end{align}
is sufficient for this purpose.
Here, we assume that \(\hat{H}_\mathrm{even}\) and \(\hat{H}_\mathrm{odd}\) consist of a sum of 2-site hermitian operators commuting with one another such that we can accurately compute the application of \(\hat{U}_\mathrm{T}\) by using the TEBD method~\cite{vidal_efficient_2003,vidal_efficient_2004,white_real-time_2004,daley_time-dependent_2004}.
As long as \(\hat{H}_\mathrm{even}\) and \(\hat{H}_\mathrm{odd}\) respect the Abelian symmetry, the application of \(\hat{U}_\mathrm{T}\) gives a symmetric basis.
The number of CPS generated by a single application of the Trotter gates is limited.
Hence, we also use \({\left[\hat{U}_\mathrm{T}{(\tau/n)}\right]}^n\) with some integer \(n\) in order to remove this limit.
The parameter \(\tau \) is optimally chosen so that the truncation errors are not severe.

\subsection{Compatibility with the hybrid approach}

Recently, the hybrid approach of the purification and the sampling has been proposed~\cite{chung_minimally_2019,chen_hybrid_2020}.
In this approach, the local Hilbert space of some sites is enlarged likewise the purification approach and the sampling is taken with respect to other sites.
In the hybrid approach, the consequent sampling of a subsystem results in larger fluctuation of projected states in comparison with that of the METTS algorithm, and thus the autocorrelation problem relaxes.
Furthermore, fluctuations of the particle number of the Bose-Hubbard model and the total magnetization of the Heisenberg model are automatically introduced even though each state in the METTS sampling has fixed values of the conserved quantities.
In other words, one can simulate the grand canonical ensemble with the Abelian symmetries respected.
A price to pay for the hybrid approach is the enlargement of local Hilbert spaces in some sites.
As the number of sites with the enlarged Hilbert space increases, the autocorrelation decreases but the computational cost for obtaining one sample increases~\cite{chung_minimally_2019}.

\begin{figure}
  \centering
  \includegraphics[width=0.95\linewidth]{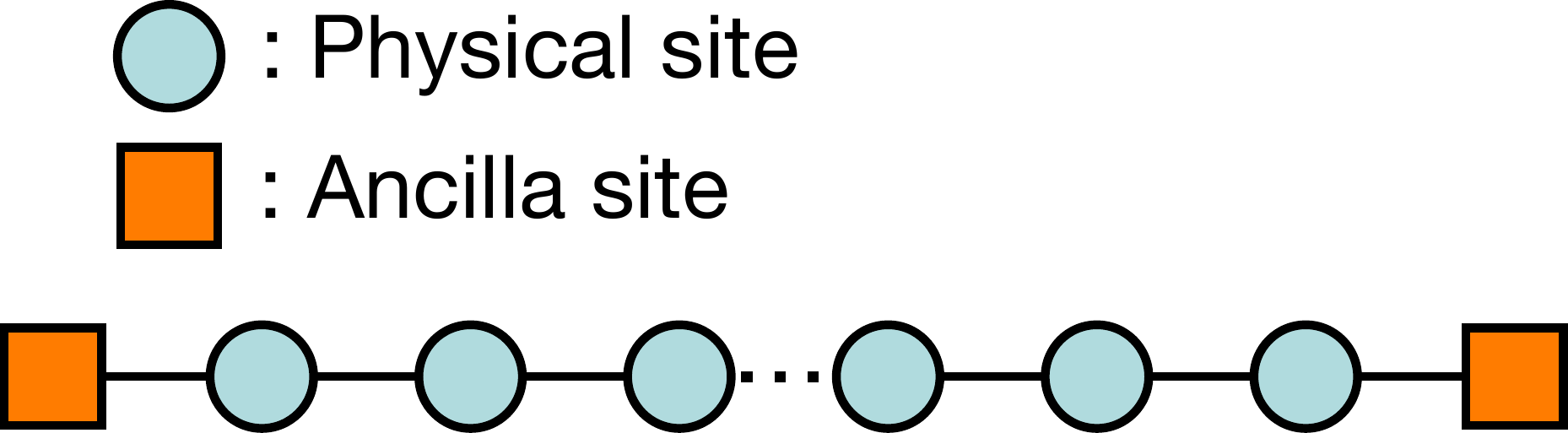}
  \caption{The alignment of physical sites (blue circles) and ancilla sites (orange squares) which does not require any additional changes of operators for enlarging local Hilbert spaces of edge sites.}\label{fig:align}
\end{figure}

Combining the hybrid approach with the above-mentioned procedure for reducing the autocorrelation problem, one can reduce the price and enjoy the access to the grand canonical ensemble.
Specifically, we enlarge the local Hilbert space of the two edge sites.
By separating the enlarged space into physical and ancilla sites, we can represent the \(L\)--site system  as a \((L+2)\)--site system shown in Fig.~\ref{fig:align}.
With the alignment in Fig.~\ref{fig:align}, one can apply any operators to physical sites in the same way as in a system without ancilla sites.
The ancilla sites do not introduce unnecessary entanglement or long-range interactions unlike other alignments.
With only two enlarged sites, the decrease of the autocorrelation by the hybrid approach is limited~\cite{chung_minimally_2019} but one can decrease the autocorrelation further by the applications of the Trotter gates.

In the next section, we present the performance tests of our approach.
For the tests, we use the \(L\)-site 1D Bose-Hubbard model at unit filling,
\begin{align}
  \label{eq:BoseHubbard}
  \hat{H} = -J\sum^{L-1}_{m=1}\left(\hat{b}^\dagger_m \hat{b}_{m+1} + \mathrm{H.c.}\right) +\frac{U}{2}\sum^L_{m=1}\hat{n}_m(\hat{n}_m - 1).
\end{align}
Here, \(J\) is the hopping integral, \(U\) is the on-site Hubbard interaction, 
\(\hat{b}_m\) (\(\hat{b}^\dagger_m\)) annihilates (creates) a boson at site \(m\), and \(\hat{n}_m = \hat{b}^\dagger_m \hat{b}_m\). 
We set \(L\) to be even.
When we simulate the grand canonical ensemble by the hybrid approach, we replace \(\hat{H}\) to \(\hat{H}-\mu \sum_m \hat{b}^\dagger_m \hat{b}_m\) with the chemical potential \(\mu \).
For \(\hat{H}_\mathrm{even}\) and \(\hat{H}_\mathrm{odd}\), we define
\begin{align}
  \begin{aligned}
  \hat{H}_\mathrm{even} = &-J\sum^{L/2-1}_{m=1}\left(\hat{b}^\dagger_{2m}\hat{b}_{2m+1} + \mathrm{H.c.}\right) \\
  & + \sum^{L-1}_{m=2} \frac{U^\prime}{4}\hat{n}_{m}(\hat{n}_{m} - 1)
  \end{aligned}
\end{align}
and
\begin{align}
  \begin{aligned}
  \hat{H}_\mathrm{odd} = &-J\sum^{L/2}_{m=1}\left(\hat{b}^\dagger_{2m-1}\hat{b}_{2m} + \mathrm{H.c.}\right) \\ 
  & + \sum^{L}_{m=1} \frac{U^\prime}{4}(1+\delta_{m,1}+\delta_{m, L})\hat{n}_{m}(\hat{n}_{m} - 1),
  \end{aligned}
\end{align}
where \(\delta_{i,j}\) is the Kronecker delta and \(U^\prime \) is either \(U\) or 0.
We project an imaginary-time evolved state into a CPS \(\ket{i}\) for even steps of the MCMC sampling and into a symmetric base \({\left[\hat{U}_\mathrm{T}{(\tau/n)}\right]}^n\ket{i}\) for odd steps.

\section{Performance tests\label{sec:results}}
\subsection{Analysis with the second largest magnitude eigenvalue}
For large lag \(t\), the autocorrelation   
\begin{align}
    C(t) = \frac{1}{M-t}\sum^{M-t}_i X_i X_{i+t} - {\left(\frac{1}{M}\sum^M_i X_i\right )}^2
\end{align}
is expected to decay exponentially~\cite{sokal_monte_1997} as
\begin{align}
    C(t) \approx C(0) \mathrm{e}^{-\frac{t}{\tau_\mathrm{exp}}}.
\end{align}
Here, \(M\) is the number of samples and \(X_i\) is some observed value of sample \(i\).
The exponential autocorrelation time \(\tau_\mathrm{exp}\) is bounded by the slowest relaxation mode of the Markov chain which is determined by the second largest magnitude eigenvalue (SLME) of a transition matrix  \( \lambda_2\) as~\cite{sokal_monte_1997}
\begin{align}
\tau_\mathrm{exp} \leq -\frac{1}{\log |\lambda_2|}.
\end{align}
Since one already knows that the transition probability of our approach is given by Eq.~\eqref{eq:nmetts_prob}, 
the SLME can be obtained as long as the dimension of the Hilbert space is small for numerical diagonalization.

\begin{figure}
    \includegraphics[width=0.95\linewidth]{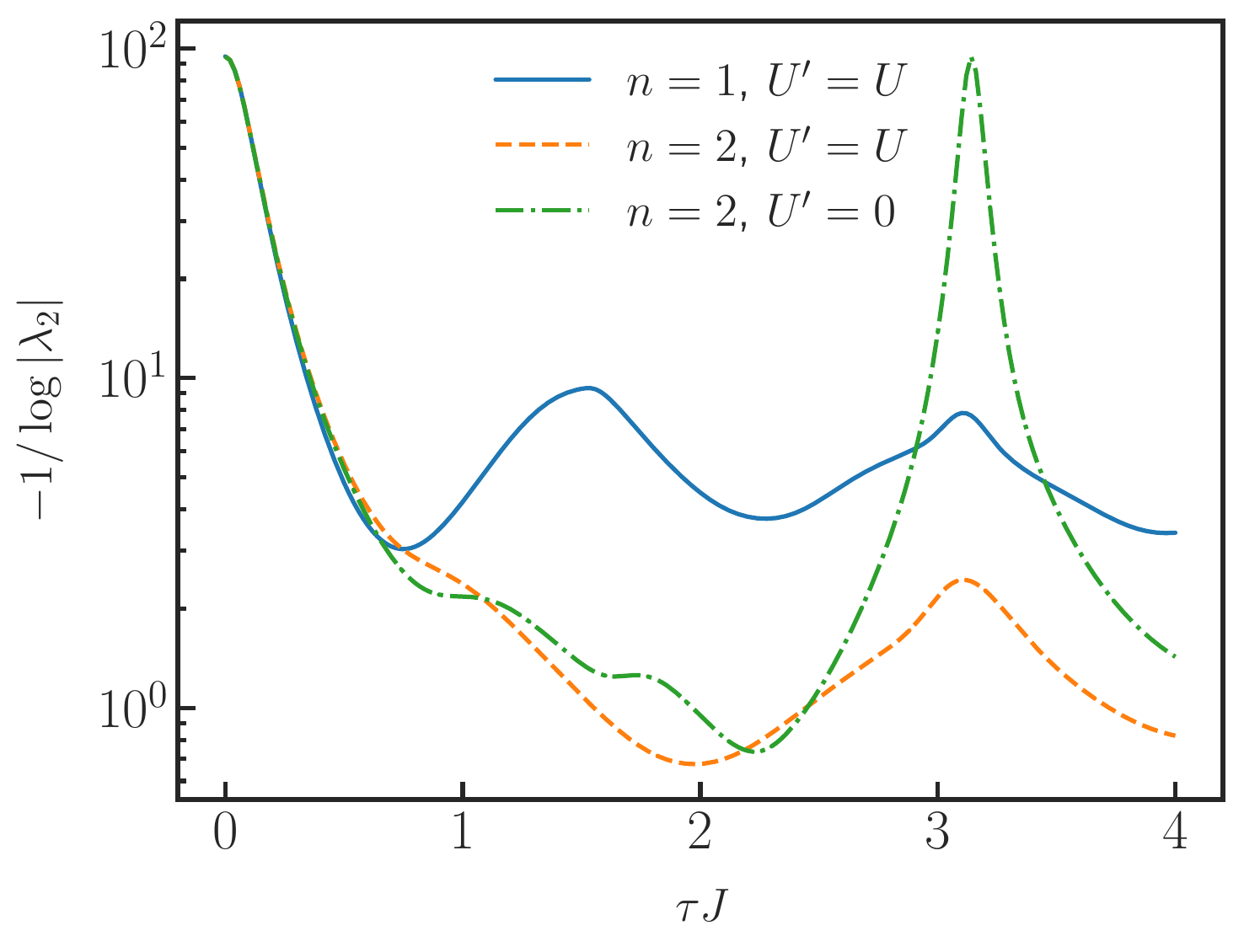}
    \caption{The \(\tau \)--dependence of the upper bound for autocorrelation time \(-1/\log|\lambda_2|\) in the 1D unit-filled Bose-Hubbard model with \(U/J = 1.0\) at inverse temperature \(\beta J = 0.25\). The system size \(L\) is set to six. For the parameters of the unitary operator \({\left[\hat{U}_\mathrm{T}{(\tau/n)}\right]}^n\), we use \((n, U^\prime) = (1, U)\) (blue solid line), \((n, U^\prime) = (2, U)\) (orange dashed line), and \((n, U^\prime) = (2, 0)\) (green dashed-dotted line).\label{fig:Bose_one}}
\end{figure}

Figure~\ref{fig:Bose_one} represents the \(\tau \)--dependence of the upper bound for autocorrelation time \( -1 / \log|\lambda_2|\) in the 1D Bose-Hubbard model of Eq.~\eqref{eq:BoseHubbard} with \(\nu=1\), \(U/J = 1.0\), and \(\beta J = 0.25\).
Here, \(\nu \) denotes the filling factor and we set the system size \(L\) to be six.
The dimension of the Hilbert space is only 462 with this setting, and thus one can construct explicitly the Hamiltonian and transition matrices.
Since the temperature of system is high, \(-1/\log|\lambda_2|\) is accordingly large in the simple METTS algorithm (\(\tau = 0\)): Around one hundred of samples can be correlated, therefore one may get an independent sample from several hundred samples.
As \(\tau \) increases, \(-1/\log|\lambda_2|\) decreases almost exponentially and reaches order of unity around \(\tau J = 1.0\). 
One can also see the limitation of a single application of the Trotter decomposition Eq.~\eqref{eq:1stTrotter} by comparing \(n=1\) and \(n=2\) data.
It should be noted that large \(\tau \) does not necessarily mean small autocorrelation as is clearly indicated in the revival behavior of the \(U^\prime = 0\) case (green dashed-dotted line).
Thus, one can significantly reduce the autocorrelation time by the application of \({\left[\hat{U}_\mathrm{T}{(\tau/n)}\right]}^n\) with \(n > 1\) and optimally chosen \(\tau \).

\begin{figure}
    \includegraphics[width=0.95\linewidth]{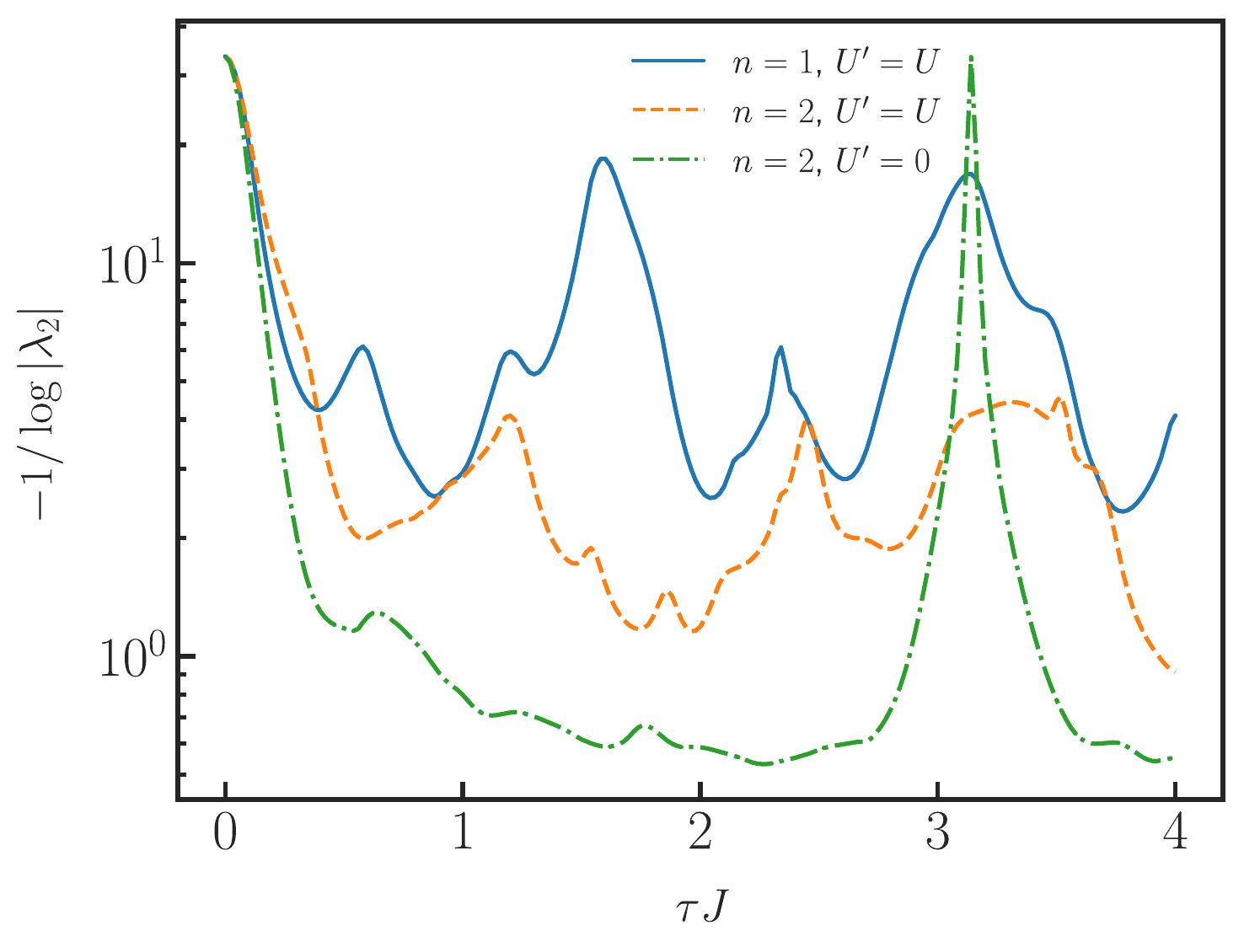}
    \caption{
        The \(\tau \)--dependence of the upper bound for autocorrelation time \(-1/\log|\lambda_2|\) in the 1D unit-filled Bose-Hubbard model with \(U/J = 20.0\). For the parameters of the unitary operator \({\left[\hat{U}_\mathrm{T}{(\tau/n)}\right]}^n\), we use \((n, U^\prime)=(1, U)\) (blue solid line), \((n, U^\prime)=(2, U)\) (orange dashed line), and \((n, U^\prime)=(2, 0)\) (green dashed-dotted line). Except the Hubbard interaction, any other parameters are taken to be the same with those in Fig.~\ref{fig:Bose_one}.\label{fig:Bose_twenty}
        }
\end{figure}

Next, we turn our attention to a system where there are some CPS being overlapped largely with some of the eigenstates of \(\hat{H}\).
Figure~\ref{fig:Bose_twenty} represents the \(\tau \)--dependence of the upper bound for autocorrelation time \(-1/\log|\lambda_2|\) at \(U/J = 20.0\).
Except the Hubbard interaction, any other parameters are taken to be the same with those of the \(U/J = 1.0\) case.
Likewise the \(U/J = 1.0\) case, the application of the Trotter gates significantly reduces the upper bound autocorrelation time.
Although the operators with \(U^\prime = U\) can reduce the autocorrelation of samples with not so large \(\tau \), 
the operator with \(U^\prime = 0\) clearly outperforms the \(U^\prime=U\) cases for small \(\tau \) and an almost uncorrelated Markov chain can be achieved around \(\tau J = 1.0\).
This result exemplifies that it is efficient to choose a Hamiltonian other than that of the system as an operator \(\hat{A}\) when some of the eigenstates of the system's Hamiltonian are well approximated as CPS.

From these SLME analyses, we see that our symmetric bases created by the application of the Trotter gates can significantly reduce the autocorrelation of samples.
In the next subsection, we show that the reduction of the autocorrelation indeed leads to the reduction of computation time.

\subsection{Computation time\label{sec:comp_time}}
\begin{figure}
    \centering
    \includegraphics[width=0.8\linewidth]{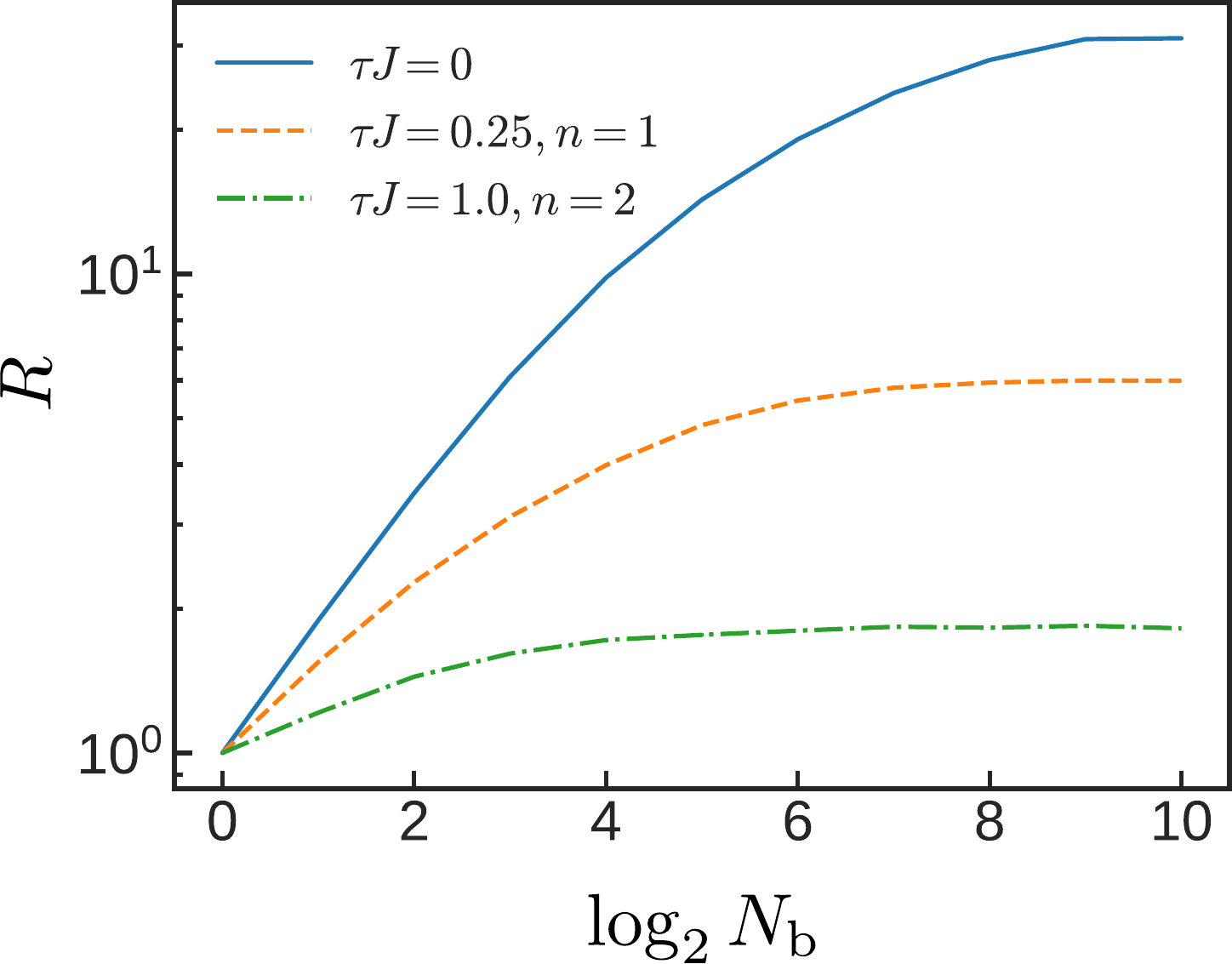}
    \caption{The ratio \(R\) as a function of the block size \(N_\mathrm{b}\) obtained from the six-site unit-filled 1D Bose-Hubbard model with \(U/J=1.0\) at inverse temperature \(\beta J = 0.25\). For the parameters of unitary operators, we use \((\tau, n, U^\prime) = (0.25J^{-1}, 1, U)\) (orange dashed line) and \((\tau, n, U^\prime) = (1.0J^{-1}, 2, U)\) (green dashed-dotted line). The solid blue line corresponds to a simulation without the Trotter gate.}\label{fig:R}
\end{figure}

The application of the Trotter gates significantly decreases the autocorrelation of samples.
However, with only this fact, we cannot conclude that this approach reduces computation time because the application of operators to a MPS is a numerically expensive task in general.
In this subsection, we show the benchmark results of MPS simulations, focusing on computation time required to obtain one uncorrelated sample.
The number of correlated samples \(R\) is estimated from the blocking analysis~\cite{gubernatis_quantum_2016} as \(R = \sigma^2_\mathrm{b}/\sigma^2 \),
where \(\sigma \) is the standard error of the total energy calculated from bare correlated samples,
\begin{widetext}
\begin{align}
    \sigma = \frac{1}{\sqrt{M}}\sqrt{\frac{1}{M}\sum^M_{i=1}{\left(\braket{\phi(i)|\hat{H}|\phi(i)} - \frac{1}{M}\sum^M_{j=1} \braket{\phi(j)|\hat{H}|\phi(j)}\right)}^2},
\end{align}
and \(\sigma_\mathrm{b}\) is the standard error calculated from blocked uncorrelated samples,
\begin{align}
    \sigma_\mathrm{b} = \sqrt{\frac{N_\mathrm{b}}{M}}\sqrt{\frac{N_\mathrm{b}}{M}\sum^{M/N_\mathrm{b}}_{i=1}{\left(\frac{1}{N_\mathrm{b}}\sum^{N_\mathrm{b}}_{j=1} \braket{\phi(N_\mathrm{b}(i-1)+j)|\hat{H}|\phi(N_\mathrm{b}(i-1)+j)} - \frac{1}{M}\sum^M_{j=1} \braket{\phi(j)|\hat{H}|\phi(j)}\right)}^2},
\end{align}
\end{widetext}
with sufficiently large block size \(N_\mathrm{b}\).
As \(N_\mathrm{b}\) increases, the ratio \(R\) also increases and approaches to a saturated value as shown in Fig.~\ref{fig:R}.
This saturated \(R\) corresponds to the number of correlated successive samples.
From \(R\) at the saturation and the averaged time elapsed to obtain one sample \(t_\mathrm{samp}\), we define the time required to obtain one \textit{uncorrelated} sample \(t_\mathrm{unc}\) as \(t_\mathrm{unc} \equiv R t_\mathrm{samp}\).
We set the maximum occupation number of boson per site to six and the truncation error to \(10^{-10}\).
The initial CPS is a classical unit-filled Mott state \(\prod_{i} b^\dagger_i \ket{0}\) where \(\ket{0}\) is the vacuum state.
For the imaginary-time evolution of MPS, we use the TEBD method~\cite{vidal_efficient_2003,vidal_efficient_2004,daley_time-dependent_2004} with the optimized Forest-Ruth-like decomposition~\cite{omelyan_optimized_2002}.
We set the imaginary time step to 0.0625\(J^{-1}\).
We perform all the simulations in this subsection on a single thread of Intel Xeon E5--2683 v4 processor and use the same pseudo-random number sequence.

\begin{table*}
    \caption{Performance tests of simulations for the 1D Bose-Hubbard model with \(L=6\), \(\nu = 1\), \(U/J = 1.0\), and \(\beta J=0.25\). Here, \(\tau \), \(n\), and \(U^\prime \) are the parameters which determine the unitary operator \({\left[\hat{U}_\mathrm{T}{(\tau/n)}\right]}^n\). \(R\) is the square of the ratio of the standard errors of blocked uncorrelated and bare correlated samples, \(\sigma^2_\mathrm{b}/\sigma^2\), at the saturation with respect to \(N_\mathrm{b}\) and indicates the effective number of correlated samples. \(t_\mathrm{samp}\) is the averaged time for obtaining one sample. \(t_\mathrm{unc}\) is the estimation of the time required to obtain one uncorrelated sample. \(1\sigma \) is the one-sigma uncertainty of \(\braket{\hat{H}}\), and \(N_\mathrm{samp}\) is the number of samples used to estimate the average and the one-sigma uncertainty of \(\braket{\hat{H}}\). The thermal expectation value obtained by the exact diagonalization is \(\braket{\hat{H}}/J = -0.9373\).\label{table:six}}
    \begin{ruledtabular}
        \begin{tabular}{ccccccccc}
            \(\tau \) (\(1/J\)) & \(n\) & \(U^\prime \) &\(R\) & \(t_\mathrm{samp}\) (s) & \(t_\mathrm{unc}\) (s) & \(\braket{\hat{H}}/J\) & 1\(\sigma/J\) &\(N_\mathrm{samp}\) \\
            \hline
            0 &  & & 38.9 & \(4.57 \times 10^{-2}\) & 1.78 & -0.9550 &\( 1.2 \times 10^{-2}\) & 1048576\\
            0.25 & 1 & \(U\) & 5.99 & \(6.39 \times 10^{-2}\) & 0.383 & -0.9400 &\( 4.5 \times 10^{-3}\) & 1048576 \\
            1.0 & 2 & \(U\) & 1.84 & \(8.96 \times 10^{-2}\) & 0.165 & -0.9382 & \( 2.4 \times 10^{-3}\) & 1048576 \\
        \end{tabular}
    \end{ruledtabular}
\end{table*}

At first, we simulate the 1D Bose Hubbard model with \(L=6\), \(\nu=1\), \(U/J = 1.0\), and \(\beta J = 0.25\) in order to confirm the results of the SLME analysis summarized in Fig.~\ref{fig:Bose_one} and compare the thermal expectation value of \(\hat{H}\) obtained by the METTS algorithm to that obtained by the exact diagonalization.
Table~\ref{table:six} shows the summary of the simulations.
In order to obtain precise numerical data, we sample much more states than the dimension of the entire Hilbert space.
The order of the estimated number of correlated samples \(R\) is consistent with the SLME analysis in Fig.~\ref{fig:Bose_one}: Several tens of samples are correlated in the simulation without the Trotter gates and there exists little autocorrelation in the simulation with the Trotter gates with \(\tau J = 1.0\). 
Consequently, \(t_\mathrm{unc}\) in the simulation with the Trotter gates is roughly one tenth of that in the simulation without the Trotter gates.
The values of \(\braket{\hat{H}}/J\) in the simulations with the Trotter gates agree with the exact value within one-sigma uncertainty.

\begin{table*}
    \caption{Performance tests of simulations for 1D Bose-Hubbard model with \(L=50\), \(\nu=1\), \(U/J = 1.0\), and \(\beta J = 0.25\). The symbol \(\geq \) means that \(R\) does not converge with \(N_\mathrm{samp}\) samples and we can estimate only a lower bound. See the caption of Table~\ref{table:six} for the definition of the other symbols.\label{table:one}}
    \begin{ruledtabular}
        \begin{tabular}{ccccccccc}
            \(\tau \) (\(1/J\)) & \(n\) & \(U^\prime \) &\(R\) & \(t_\mathrm{samp}\) (s) & \(t_\mathrm{unc}\) (s) & \(\braket{\hat{H}}/J\) & 1\(\sigma/J\) &\(N_\mathrm{samp}\) \\
            \hline
            0 &  & & \(\geq \) 78.4 & 0.741 & \(\geq \) 58.1 & -11.461 &\( \geq 9.7 \times 10^{-2}\) & 262144\\
            1.0 & 2 & \(U\) & 3.52 & 4.96 & 17.5 & -11.500 & \( 5.4 \times 10^{-2}\) & 32768 \\
            1.0 & 2 & 0 & 4.12 & 4.49 & 18.5 & -11.512 &\( 5.0 \times 10^{-2}\) & 32768 \\
        \end{tabular}
    \end{ruledtabular}
\end{table*}

Next, in order to demonstrate that our approach is efficient in large systems, we perform the same simulations with \(L=50\).
Table~\ref{table:one} shows the performance of simulations for the 1D Bose-Hubbard model with \(\nu=1\), \(U/J = 1.0\), and \(\beta J = 0.25\).
The values of \(\braket{\hat{H}}/J\) computed by the three METTS simulations agree within one-sigma uncertainty. 
This result confirms that the use of the symmetric basis created by the operation of the Trotter gates does not affect the stationary distribution.
As expected, the application of the Trotter gates increases \(t_\mathrm{samp}\) roughly by six times.
On the contrary, it substantially decreases the effective number of correlated samples \(R\) by at least one twentieth.
Consequently, \(t_\mathrm{unc}\) reduces roughly to one third of the value of the METTS algorithm without the Trotter gates.
Thus, we can conclude that the application of the Trotter gates noticeably reduces the computation time of the METTS algorithm.
Comparing the case of \(U^\prime = U\) with that of \(U^\prime = 0\) in Table~\ref{table:one}, there is no significant difference.
The absence of the difference is consistent with the SLME analysis summarized in Fig.~\ref{fig:Bose_one}.

\begin{table*}
    \caption{Performance tests of simulations for the 1D Bose-Hubbard model with \(L=50\), \(\nu=1\), \(U/J = 20.0\), and \(\beta J=0.25\). See the caption of Tables~\ref{table:six} and~\ref{table:one} for the definitions of symbols.\label{table:twenty}}
    \begin{ruledtabular}
        \begin{tabular}{ccccccccc}
            \(\tau \) (\(1/J\)) & \(n\) & \(U^\prime \)  &\(R\) & \(t_\mathrm{samp}\) (s) & \(t_\mathrm{unc}\) (s) & \(\braket{\hat{H}}/J\) & \(1\sigma / J\) &\(N_\mathrm{samp}\) \\
            \hline
            0 &  & & \(\geq \) 552 & 0.764 & \(\geq \) 422 & 55.62 &\(\geq 1.00 \) & 262144\\
            1.0 & 2 & \(U\) & 28.1 & 2.66 & 74.7 & 54.70 & \( 0.60 \) & 32768 \\
            1.0 & 2 & 0 & 8.49 & 2.80 & 23.8 & 53.98 & \(0.41\) & 16384 \\
        \end{tabular}
    \end{ruledtabular}
\end{table*}

We also investigate the performance of the METTS algorithm with the Trotter gates for the 1D Bose-Hubbard model with \(L=50\), \(\nu=1\), \(U/J = 20.0\), and \(\beta J = 0.25\), where there are some CPS being overlapped largely with some of the eigenstates of the system's Hamiltonian.
Table~\ref{table:twenty} shows the performance tests of simulations for this case.
With this setting, the effective number of correlated sample \(R\) in the METTS algorithm without the Trotter gates is larger than five hundred.
It should be noticed that since this value is only a lower bound estimated from 262144 samples, it is possible that the true number of correlated samples is much larger.
The application of the Trotter gates reduces such a very large \(R\) to 28.1 with \(U^\prime = U\) and to only 8.49 with \(U^\prime = 0\), which means a tough autocorrelation problem can be significantly relaxed with the Abelian symmetry respected.
One can also confirm the superiority of the Trotter gates with \(U^\prime = 0\), which is consistent with the SLME analysis summarized in Fig.~\ref{fig:Bose_twenty}.
Although \(t_\mathrm{samp}\) increases roughly by four times due to the application of the Trotter gates with \(U^\prime = 0\), 
\(t_\mathrm{unc}\) reduces roughly by one eighteenth.
This improvement means that error bars in the METTS algorithm with the Trotter gates with \(\hat{U}^\prime = 0\) are about four time smaller than those in the ordinary METTS algorithm with the same computation time.

\subsection{Hybrid approach}
In addition to the canonical ensemble, we check the performance of the combination of the application of the Trotter gates and the hybrid approach for the simulations of the grand canonical ensemble.
We take the strong \(U\) limit and treat bosons as hard-core bosons which do not occupy the same site.
A strong advantage of taking the hardcore boson limit is that the system can be mapped on to free fermions~\cite{girardeau_relationship_1960}, where exact diagonalization with a large system is feasible.
Hence, we can carry out the performance tests of the METTS algorithm on the basis of the comparison with the results obtained by using the exact diagonalization.

\begin{table*}
    \caption{Performance tests of simulations for the 1D Bose-Hubbard model with \(L=50\) and \(\beta J = 5.0\) in the strong \(U\) limit. The chemical potential \(\mu \) is set to be \(-2.0J\). Here, \(\kappa \) is the compressibility, \(1\sigma \) is the one-sigma uncertainty estimated from the jackknife analysis. The exact numerical value of the compressibility \(\kappa \) is \(11.866 J^{-1}\). See the caption of Tables~\ref{table:six} for the definitions of the other symbols.\label{table:hardcore}}
    \begin{ruledtabular}
        \begin{tabular}{cccccccc}
            \(\tau \) (\(1/J\)) & \(n\) &\(R\) & \(t_\mathrm{samp}\) (s) & \(t_\mathrm{unc}\) (s) & \(\kappa J\) & \(1\sigma J\) &\(N_\mathrm{samp}\) \\
            \hline
            0 &  & 224.6 & 1.72 & 386 & 11.85 & 0.74  & 65536\\
            3.6 & 2  & 68.3 & 2.04 & 139 & 11.60 & 0.47 & 65536 
        \end{tabular}
    \end{ruledtabular}
\end{table*}

In the grand canonical ensemble, one can obtain the fluctuations of the total number of particles which is related to the compressibility \(\kappa \) as
\begin{align}
    \kappa = \frac{\partial \braket{\sum_i \hat{n}_i }}{\partial \mu} = \beta \left(\Braket{{\left(\sum_i \hat{n}_i\right)}^2} - \Braket{\sum_i \hat{n}_i}^2\right).~\label{eq:compressibility}
\end{align}
Table~\ref{table:hardcore} shows the performance tests of the simulations of the grand canonical ensemble where we take the 1D Bose-Hubbard model in the strong \(U\) limit at \(L=50\), \(\mu/J = -2.0\), and \(\beta J = 5.0\).
These simulations are also performed on a single thread of Intel Xeon E5--2683 v4 processor and use the same pseudo-random number sequence, and we use the second order Suzuki-Trotter decomposition with the time step 0.025\(J^{-1}\) for imaginary-time evolution.
We estimate the uncertainty of \(\kappa\) from the jackknife analysis~\cite{gubernatis_quantum_2016}.
The estimated values of \(\kappa \) are consistent with the exact value within the one-sigma error, thus confirming that the grand canonical ensemble is simulated properly.
Moreover, the computation time required for one uncorrelated data \(t_\mathrm{unc}\) is reduced roughly by one-third by thanks to the Trotter gates.
In short, our approach based on the application of the Trotter gates is effective in the hybrid approach.

Let us compare the efficiency of the simulations of the grand canonical ensemble with that of the canonical ensemble.
Specifically, we simulate the canonical ensemble of the 1D Bose-Hubbard model in the strong \(U\) limit at \(L=50\), \(\nu = 0.08\), and \(\beta J = 5.0\) by using the METTS algorithm with the Trotter gates.
For the comparison, the filling factor \(\nu \) is determined to be close to that of the grand canonical ensemble simulated in Table~\ref{table:hardcore}, \(\nu = 0.074\), and we use the same Trotter gates.
In the canonical ensemble, the estimated number of correlated samples \(R\) is 9.2, which is around one-seventh of the value in the grand canonical ensemble 68.3.
Therefore, the canonical ensemble is preferred unless one wants to compute observables converted from the fluctuations of conserved quantities, such as the compressibility and the magnetic susceptibility.
The inefficiency of the grand canonical ensemble can be attributed to the fact that while the sampling space is much larger than that in the canonical ensemble, the total particle number can change only by at most two in one Monte Carlo step in our setting, where there are only two ancilla sites.
Although increasing ancilla sites reduces the number of correlated samples but also increases the numerical cost of obtaining one sample~\cite{chung_minimally_2019}. 

\begin{figure}
    \includegraphics[width=\linewidth]{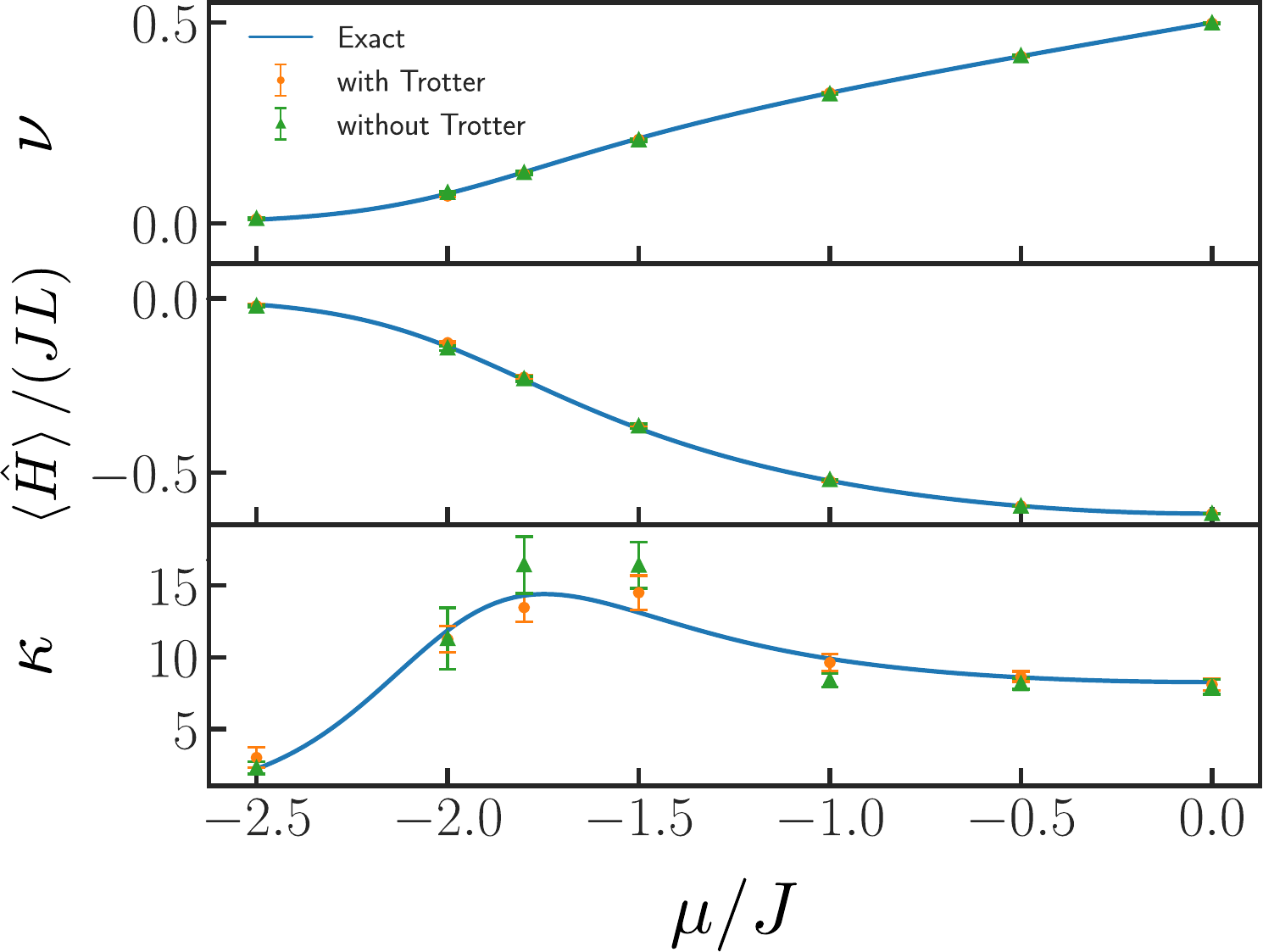}
    \caption{The chemical potential \(\mu \)--dependencies of the filling factor \(\nu \) (Upper panel), the internal energy per site \(\braket{\hat{H}}/(JL)\) (Middle panel), and the compressibility \(\kappa \) (Lower panel) of the 1D Bose-Hubbard model in the large \(U\) limit calculated by using the METTS algorithms with and without the Trotter gates. The inverse temperature \(\beta \) is \(5.0J^{-1}\). For the parameters characterizing the Trotter gates, we use \(\tau J = 3.6\) and \(n=2\). The expectation values and the one-sigma error bars of the METTS results are estimated from successive 8192 samples.\label{fig:mu_dep}}
\end{figure}

Figure~\ref{fig:mu_dep} represents the chemical potential \(\mu \)--dependencies of the filling factor \(\nu \), the internal energy per site \(\braket{\hat{H}}/(JL)\), and the compressibility \(\kappa \) obtained by using the METTS algorithm with and without the application of the Trotter gates.
The expectation values and the one-sigma error bars are estimated from successive 8196 samples.
For the filling factor and the internal energy, both of the METTS algorithms with and without the Trotter gates give sufficiently precise and accurate values for the entire region of the chemical potential.
On the contrary, as for the compressibility, the deviation from the results obtained by using the exact diagonalization is visible in both algorithms.
Nevertheless, the error bars of the METTS algorithm with the Trotter gates are smaller than those of the METTS algorithm without the Trotter gates, especially at \(\mu/J = -1.8\), where the compressibility takes a maximum value.
This result indicates that the application of the Trotter gates to the METTS algorithm allows for a more efficient description of the grand canonical ensemble.

\section{Summaries\label{sec:summaries}}
We improved the minimally entangled typical thermal states (METTS) algorithms by adding the operation of a series of Trotter gates which transforms the symmetric basis~\cite{binder_symmetric_2017}.
We performed the analysis using the second largest magnitude eigenvalue of a transition matrix for the one-dimensional Bose-Hubbard model with unit filling in order to show that a correlation of successive samples significantly decreases by applying the Trotter gates. 
From the performance tests for the same model, we confirmed that the reduction of the autocorrelation leads to the reduction of computation time and thus improves the numerical efficiency of the METTS algorithm. 
We showed that the application of the Trotter gates can be combined with the recently proposed hybrid approach~\cite{chung_minimally_2019,chen_hybrid_2020} and improves the efficiency of simulations of the grand canonical ensemble.
Therefore, the improved approach relaxes the autocorrelation problem of the METTS algorithm without breaking the Abelian symmetries.
The improved METTS algorithm is applicable potentially to many problems at finite temperatures, such as transport, quench dynamics, and the magnetization curve.

\begin{acknowledgments}
The MPS calculations in this work are performed with ITensor library, http://itensor.org.
This work was financially supported by KAKENHI from Japan Society for Promotion of Science: Grant No.\ 18K03492, No.\ 18H05228, and No.\ 20K14377, by CREST, JST No.\ JPMJCR1673, and by MEXT Q-LEAP Grant No.\ JPMXS0118069021.
\end{acknowledgments}
\bibliography{../Library}
\end{document}